\documentclass[12pt]{article}
\usepackage[dvips]{graphics}
\begin{document}
\def\B{{\mathcal B}}                        \def\bea{\begin{eqnarray}}
\def\C{{\mathcal C}}                        \def\eea{\end{eqnarray}}
\def\D{{\mathcal D}}                        \def\ig{\includegraphics}
\def\L{{\mathcal L}}
\def\M{{\mathcal M}}
\def\Bg{{\mathcal B}_g}
\def\Cr{{\mathcal C}_r}
\def\Cy{{\mathcal C}_y}
\def\Dr{{\mathcal D}_r} 
\def\VgB{V_g^{\mathcal B}}
\def\Drg{{\mathcal D}_{rg}}                     
\def\QgBr{Q_g^{\mathcal B}(r)}
\def\RgBr{R_g^{\mathcal B}(r)}
\def\SgBr{S_g^{\mathcal B}(r)}              
\def\QgBrl{Q_g^{\mathcal B}[r(l)]}
\def\RgBrl{R_g^{\mathcal B}[r(l)]}                    
\def\SgBrl{S_g^{\mathcal B}[r(l)]}                  
\def\BUBg{{\mathcal B}\cap {\mathcal B}_g}  
\def\BUCr{{\mathcal B}\cap {\mathcal C}_r}
\def\BgUCr{{\mathcal B}_g\cap {\mathcal C}_r}  
\def\PgBl{{\mathcal P}_g^{\mathcal B}(l)}
\def\BUBgUCr{{\mathcal B}\cap {\mathcal B}_g\cap {\mathcal C}_r}
\def\PscBl{{\mathcal P}_{sc}^{\mathcal B}(l)}
\def\PdBl{{\mathcal P}_1^{\mathcal B}(l)}
\def\PsBl{{\mathcal P}_3^{\mathcal B}(l)}
\def\vfGBl{\varphi_{\Gamma}^{\mathcal B}(l)}  
\def\vfuBl{\varphi_{un}^{\mathcal B}(l)}

\title{\Large \bf Cosmic crystallography: the hyperbolic isometries \\ }
\author{A.F.F. Teixeira \thanks{teixeira@cbpf.br} \\ 
{\small Centro Brasileiro de Pesquisas F\'{\i}sicas }\\
{\small 22290-180 Rio de Janeiro-RJ, Brazil} \\
       }
\date{\today}
\maketitle
\begin{abstract}									
All orientation preserving isometries of the hyperbolic three-space are studied, 
and the probability density of conjugate pair separations for each isometry is 
presented. The study is relevant for the cosmic crystallography, and is the 
theoretical counterpart of the mean histograms arising from computer simulations 
of the isometries. 
\end{abstract}
\section{Introduction}								
\setcounter{equation}{0} 
Cosmic crystallography is a method to help finding  the geometry and 
the topology of the universe \cite{lelalu}. 
In a close analysis of the method, a description has been presented of how 
each isometry of the universe gives its individual contribution 
to a pair separation histogram (PSH) \cite{spikes1}. 
More recently, the isometries of the infinite 3D {\it euclidean} space 
were studied in some detail, and the expected (theoretical) individual 
contribution of each isometry to the PSH was described \cite{cceuc}.

The present report investigates the orientation preserving isometries of $H^3$, 
the 3D infinite {\it hyperbolic} space with positive definite metric 
and unitary radius of negative curvature. 
To generalize for arbitrary negative curvature one simply needs dividing every 
quantity with dimension length by the radius of curvature. 
All the results obtained clearly reproduce their euclidean counterparts 
when the radius of curvature tends to infinity.

\section{The probability density of conjugate pair separations}   
In $H^3$ we assume a spherical solid ball $\B$. 
Under an isometry $g$ of $H^3$ the ball occupies a new position $\Bg$; we only 
consider isometries such that the balls $\B$ and $\Bg$ intersect. 
Assuming a point $P\in\B$ and denoting as $P_g\in\Bg$ its $g$-transported, 
we call the pair $(P, P_g)$ a $g$-pair. 
We focus our attention on the $g$-pairs such that $P_g\in\BUBg$, 
and assume an infinity of points $P_g$ uniformly distributed throughout 
the intersection $\BUBg$.  
For the given isometry $g$, we ask for the probability $\PgBl dl$ that a randomly 
selected $g$-pair has hyperbolic separation lying between the values $l$ and $l+dl$; 
the probability density $\PgBl$ clearly satisfies the normalization condition 
\bea                                                       \label{1}     
\int_0^{2a}\PgBl dl = 1 ,
\eea 
where $2a$ is the diametre of the balls. 

\section{Some basic formulas}                                       
A few useful formulas of the hyperbolic trigonometry in 2D are worth having at hand. 
In a geodetic triangle with sides measuring $a, b, c$ and corresponding opposite 
angles measuring $\alpha, \beta, \gamma$, we have 
\begin{itemize}
\item the law of sines 
\bea                                                        \label{2}
\frac{\sin{\alpha}}{\sinh{a}}=\frac{\sin{\beta}}{\sinh{b}}=
                                          \frac{\sin{\gamma}}{\sinh{c}}; 
\eea 
\item the first law of cosines (cyclic)
\bea                                                        \label{3}
\cosh{a}=\cosh{b}\cosh{c}-\sinh{b}\sinh{c}\cos{\alpha} ; 
\eea 
\item the second law of cosines (cyclic)
\bea                                                        \label{4}
\cos{\alpha}=-\left(\cos{\beta}\cos{\gamma}-\sin{\beta}\sin{\gamma}\cosh{a}\right) . 
\eea
\end{itemize}
We also need a few 2D relations between lengths of arcs of geodesics, horocycles 
and equidistant curves to a geodesic; see figure 1 for visualization: 
\bea                                                         \label{5} 
e'=g\cosh{r}, \hskip5mm h'=h\cosh{r}, \hskip5mm h=2\sinh{(g/2)}, \\ 
\nonumber e^f=\cosh(g/2), \hskip18mm \tanh p'=\tanh p\cosh q, 
\eea 
and as a consequence 
\bea                                                       	\label{6}	
\sinh{(g'/2)}=\sinh{(g/2)}\cosh{r} \, .
\eea

\vskip3mm   \hskip3cm\scalebox{0.5}{\ig{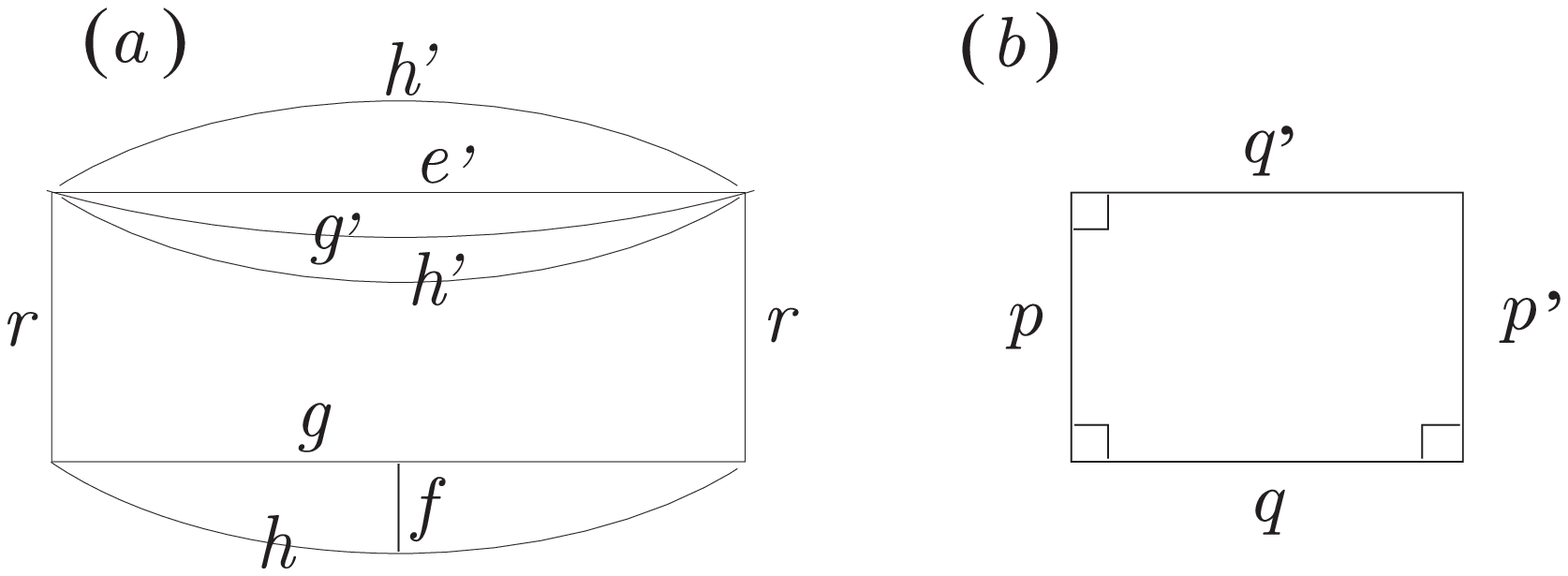}}    \vskip3mm    
    
\noindent {\small {\bf Figure 1} Some geometric objects in the plane hyperbolic 
geometry. \\
$(a)$ $g, g'$, and $r$ are geodetic arcs; 
$h$ and $h'$ are arcs of horocycles; $e'$ is an arc equidistant to $g$; 
the arcs $r$ are perpendicular to $g$ and to $e'$; 
the geodetic arc $f$ is orthogonal to both $g$ and $h$. \\
$(b)$ a geodetic quadrilateral with three right angles, assuming  
$\sinh p\sinh q<1$.}

\vskip5mm 
We often use the line element of $H^3$ in the cylindrical coordinates, 
\bea                                                              \label{7} 
dl^2=d\rho^2+\sinh^2\!\rho\, d\phi^2+\cosh^2\!\rho\, d\zeta^2 ;
\eea 
\noindent we also use 
\bea                                          				\label{8}      
dl^2=-dW^2+dX^2+dY^2+dZ^2 \, , \hskip5mm W=\sqrt{1+X^2+Y^2+Z^2} \, , 
\eea 
where 
\bea              								\label{9} 
W=\cosh\rho\cosh\zeta, \hskip3mm X=\sinh\rho\cos\phi, \hskip3mm 
Y=\sinh\rho\sin\phi, \hskip3mm Z=\cosh\rho\sinh\zeta \hskip2mm . 
\eea 
\noindent In these coordinates the separation $l$ between 
two points $P_1, P_2$ is given by 
\bea											\label{10} 
\cosh{l} &=& W_1W_2-(X_1X_2+Y_1Y_2+Z_1Z_2) \nonumber\\ 
         &=& \cosh\rho_1\cosh\rho_2\cosh(\zeta_1-\zeta_2)-                     \sinh\rho_1\sinh\rho_2\cos(\phi_1-\phi_2) \hskip1mm .		
\eea 

For future reference, consider the special situation of the three points 
(see figure 2)
\begin{eqnarray*}											
P_1  &=  (\rho_1, 0, 0)  =  &(\cosh\rho_1; \sinh\rho_1, 0, 0) \hskip1mm , \\
Q    &=\, (0, 0, g)\,    =  &(\cosh{g}; 0, 0, \sinh{g}) \hskip1mm , \\
P_2  &=  (\rho_2, \phi, g) =  &(\cosh\rho_2\cosh{g}; \sinh\rho_2\cos\phi, 
     \sinh\rho_2\sin\phi) \hskip1mm; 
\end{eqnarray*}
the separation $l$ between the points $P_1$ and $P_2$ is then given by  
(\ref{10}), namely 
\bea											\label{11}
\cosh{l}=\cosh\rho_1\cosh\rho_2\cosh{g}-\sinh\rho_1\sinh\rho_2\cos\phi \hskip1mm .
\eea

\vskip3mm   \hskip5cm\scalebox{0.4}{\ig{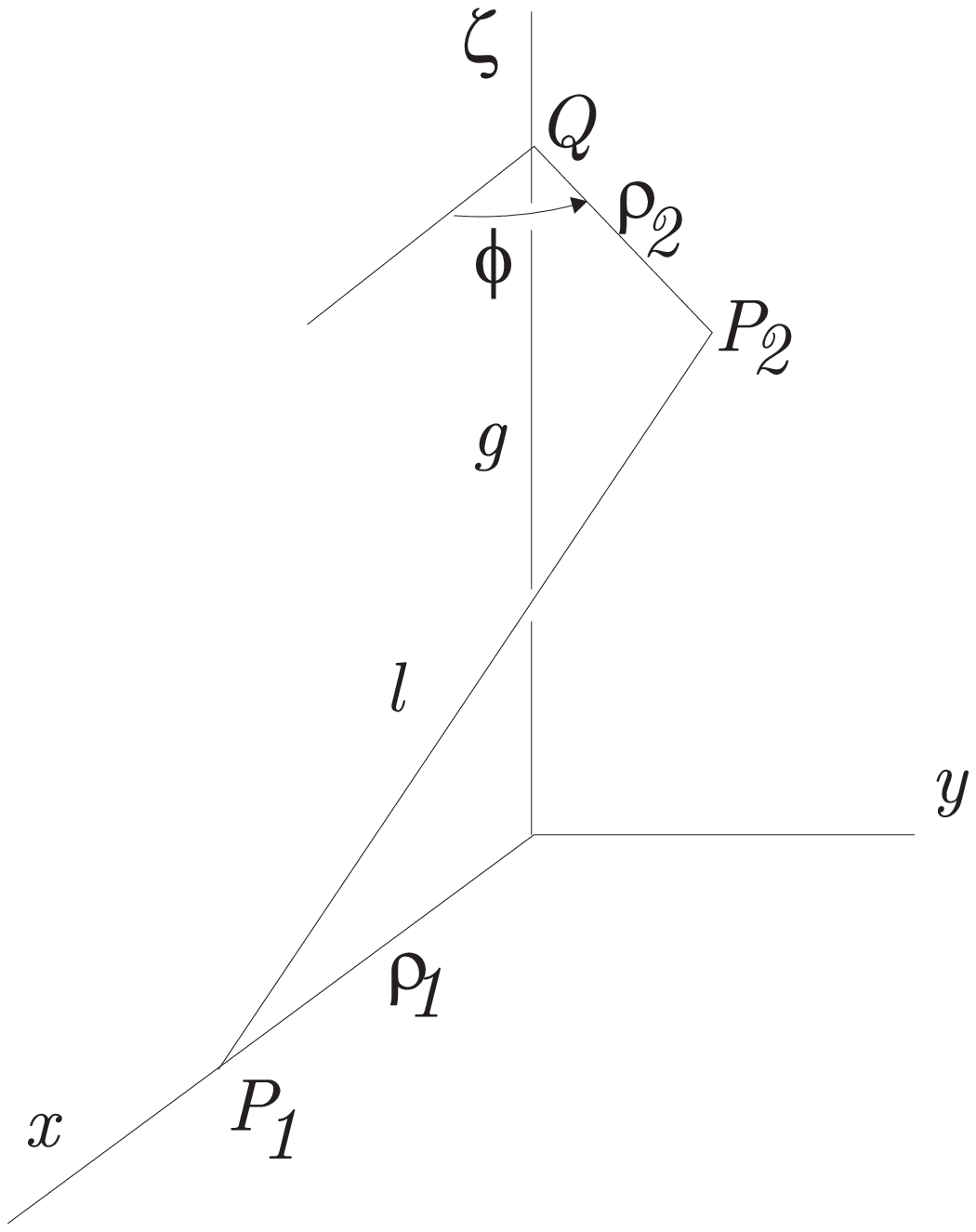}}    \vskip3mm       
    
\noindent {\small {\bf Figure 2} Relative position of points $P_1$ and $P_2$ 
with hyperbolic separation $l$. }
 
\vskip5mm 
\newpage 
\section{The isometries of $H^3$}						
We are presently interested in the isometries of $H^3$ that preserve 
the orientation. These isometries can be classified as 
\begin{itemize}
\item hyperbolic, with 5 parametres; these isometries bear some similarity with 
the 3-parametric euclidean translations; 
\item elliptic, also with 5 parametres; they are analogous to the 
also 5-parametric euclidean rotations in ordinary space; 
\item screw motions, with 6 parametres; bear some similarity with 
the also 6-parametric euclidean screw motions; 
\item parabolic, with only 4 parametres; again they remind us of  
the euclidean translations. 
\end{itemize}
\vskip1cm
In each of these isometries the intersection of the balls $\B$ and $\Bg$ is a 
rotationally symmetric solid lens, whose thickness $T$, diametre $D=2R$, 
and volume $\VgB$ we now seek. We denote as $a$ the radius of the balls, 
and $m=2M$ the separation between their centres $C$ and $C_g$; 
then we clearly have $M<a$ and 
(see figure 3) 
\bea                                                                \label{12}
T=2(a-M) \, . 
\eea 
The separation $m$ depends on the isometry $g$ one is concerned with. 
Noting that $a, M$, and $R$ make a right-angled triangle with hypotenuse $a$, 
we find 
\bea                                                                \label{13}
\cosh{a}=\cosh{M}\cosh{R} \, .
\eea 

\vskip1cm   \hskip6cm\scalebox{0.9}{\ig{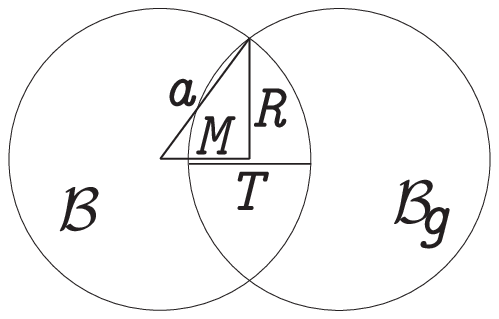}}    \vskip3mm      
   
\noindent {\small {\bf Figure 3} The balls $\B$ and $\Bg$ with radius $a$ 
intersect in a solid lens with equatorial radius $R$ and thickness $T$. 
The centres of the balls are separated $m=2M$}.
 
\vskip1cm
To have the volume $\VgB$ of the solid lens $\BUBg$
we first consider a compact cylindrical surface $\Cy$ embeded in the lens, 
and whose axis coincides with that of the lens (see figure 4);   
all points of $\Cy$ are at a fixed distance $y$ from the axis, 
so the geometry on $\Cy$ is 2D-euclidean. Denoting as $e$ the length of 
the generatrices (arcs equidistant to the geodetic axis), the area of $\Cy$ is 
\bea                                                               \label{14} 
S(y)=2\pi e\sinh{y} \, . 
\eea 

\vskip3mm  \hskip6cm\scalebox{0.7}{\ig{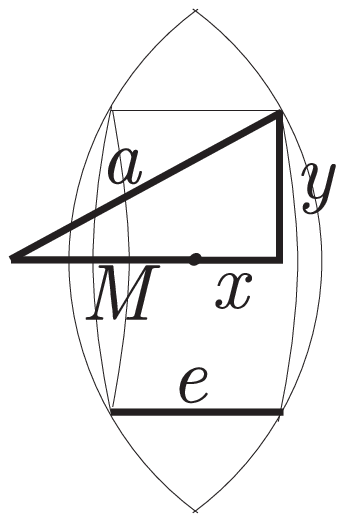}}     \vskip3mm      

\noindent {\small {\bf Figure 4} \hskip1mm Sketch of a compact cylinder $\Cy$ 
inscribed in the solid lens $\BUBg$; it has radius $y$ and generatrices $e$. }
 
\vskip5mm
\noindent We must now relate $e$ with $a, M$, and the variable radius $y$. 
In figure 4 we note a right angled triangle with sides $a$ (hypotenuse), $y$, 
and $M+x$, so we have 
\bea										      \label{15} 
\cosh{a}=\cosh{y}\cosh(M+x) \, ; 
\eea 
since from eq. (5a) we have $e/2=x\cosh{y}$ , then  
$e=2\cosh{y}\left[\cosh^{-1}\left(\frac{\cosh{a}}{\cosh{y}}\right)-M\right]$ 
\, , and 
\bea 											\label{16}	
S(y)=2\pi\sinh{2y}\left[\cosh^{-1}\left(\frac{\cosh{a}}
{\cosh{y}}\right)-M\right] \, . 
\eea 
The volume $\VgB$ of the lens is clearly 
\bea										 
\VgB&=&\int_0^{R}S(y)dy    \nonumber\\		                                   
    &=&2\pi\left[(\tanh{a}-\tanh{M})\cosh^{2}{a}-(a-M)\right] \, . \label{17}
\eea 

It can be checked that when $M=0$ we get the hyperbolic volume 
$\pi(\sinh{2a}-2a)$ of a solid ball with radius $a$, as expected. 
Also note that for small values of $a$ and $M$ we recover the euclidean 
volume $(2\pi/3)(a-M)^2(2a+M)$ of the solid lens \cite{cceuc} .

\section{Special translations}                                    
Preceding the study of the {\it general hyperbolic} isometry of $H^3$ 
we first consider the very special situation in which the axis $\L$ 
of the isometry $g$ crosses the centre $C$ of the ball $\B$. 
With $a=$ the radius of $\B$, and $t=$ the value of the translation along the 
axis, we assume $t<2a$ to have nonvanishing intersection $\BUBg$. 
In this special isometry we clearly have the equality $m=t$. 

According to eq.(\ref{6}) and figure 1, a point $P$ at a distance $r$ 
from the axis $\L$ is displaced under $t$ to a distance $l$ given by 
\bea											\label{18} 
\sinh{(l/2)}=\sinh{(t/2)}\cosh{r} \, ; 
\eea 
this is a relation involving the variable $l$ (displacement of $P$), 
the variable $r$ (distance from $P$ to the axis $\L$), 
and the parametre $t$ (the unique relevant one in this special isometry). 

We next introduce the probability $\QgBr dr$ that a randomly chosen point $P_g$ 
which is in both $\B$ and $\Bg$ be in a radial position between $r$ and $r+dr$. 
The probability density $\QgBr$ clearly is proportional to the area 
$\SgBr$ of the cylinder $\Cr$ inscribed in the solid lens $\BUBg$   
(see eq.(\ref{16})), the coefficient of proportionality being the inverse of the 
volume $\VgB$ of the lens: 
\bea											\label{19}
\QgBr=\frac{\SgBr}{\VgB}= 
\frac{2\pi}{\VgB}\sinh{2r}\left[\cosh^{-1}\left(\frac{\cosh{a}} {\cosh{r}}\right)
- \frac{t}{2}\right] \, .
\eea
The equality of the probabilities $\PgBl{dl}$ and $\QgBr{dr}$ then gives, 
using $r(l)$ obtained from (\ref{18}),   
\bea										      \label{20} \PgBl&=&\frac{dr}{dl}\QgBrl \nonumber\\
     &=&\frac{1}{\VgB}  \frac{\pi\sinh{l}}{\sinh^2t/2}
      \left[\cosh^{-1}\left(\frac{\cosh{a}\sinh{t/2}}                 
            {\sinh{l/2}}\right)-\frac{t}{2}\right] \, .                               
\eea
In the figure 5 we have four instances of $\PgBl$.  
Each plot starts abruptly on $l=t$ and vanishes when 
$\sinh(l/2)=\cosh a\tanh(t/2) $. 
They greatly differ from that of an euclidean translation, 
where $\PgBl=\delta(l-t)$, a Dirac $\delta$.

\vskip3mm  \hskip3cm\scalebox{0.6}{\ig{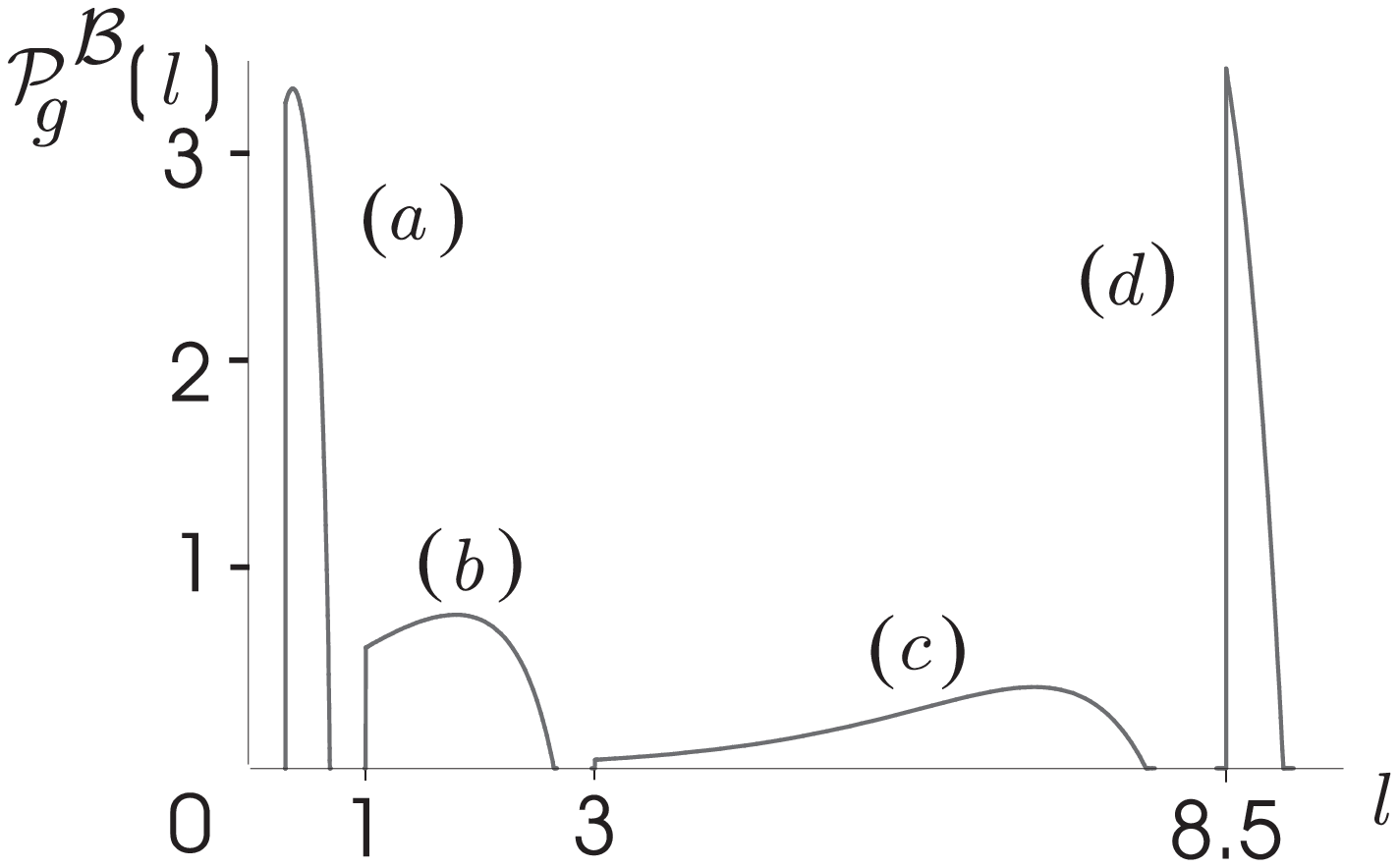}}  \vskip3mm       
    
\noindent {\small {\bf Figure 5} Probability densities $\PgBl$ of pair 
separations for hyperbolic motions $t$ of a ball $\B$ with radius $a$, 
when the axis  of the isometry crosses the centre of the ball. 
In $(a)$ we took $t=0.3$ and $a=1.5$; 
in $(b)$, $t=1$ and $a=2$;  
in $(c)$, $t=3$ and $a=4$; and   
in $(d)$, $t=8.5$ and $a=4.5$. 
Plots $(a)$ and $(d)$ loosely resemble a Dirac $\delta$ mainly because 
the corresponding solid lenses $\BUBg$ have volume $\VgB$ small in comparison 
with the volume $\pi(\sinh{2a}-2a)$ of $\B$.  
All integrated areas are unitary.}
\vskip5mm 

\newpage
\section{Special screw motions}                                      
It is very simple to generalize the probability density (\ref{20}) to further have a 
rotation $\omega$ of $\B$ around the axis $\L$ of the translation. See figure 6. 

\vskip3mm  \hskip6cm \scalebox{0.35}{\ig{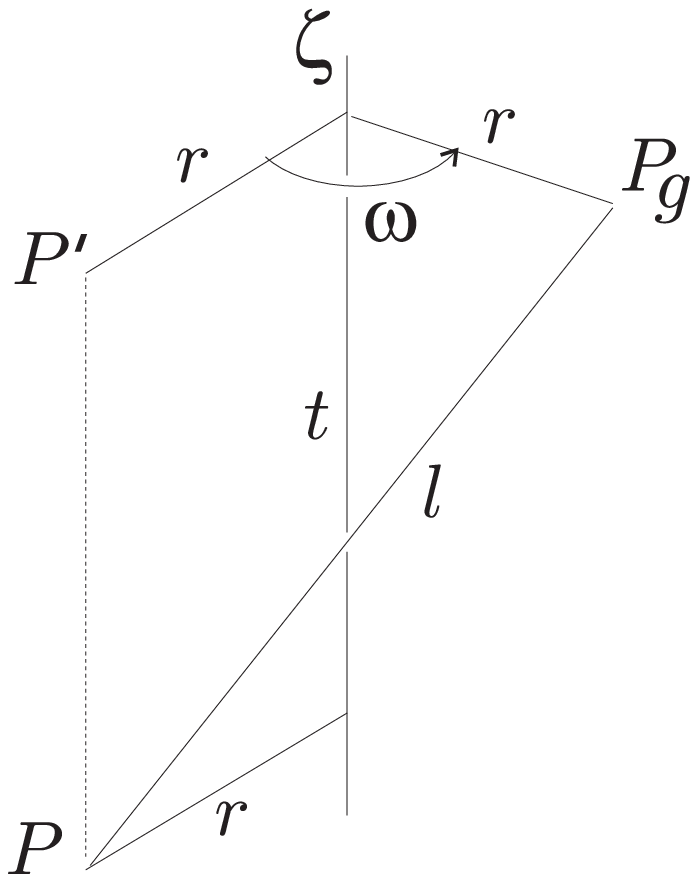}}  \vskip3mm      

\noindent {\small {\bf Figure 6} A screw motion of $H^3$ with axis along the 
$\zeta$ axis. Under the preliminary hyperbolic motion $t$ (measured on 
the $\zeta$ axis) a point $P$ is displaced to the intermediate position $P'$; 
then the rotation $\omega$ around $\zeta$ brings $P'$ to the final position 
$P_g$ geodetically separated $l$ from $P$.}  
 
\vskip5mm
The separation $l$ between a point $P$ and the corresponding $P_g$ is now given 
by (\ref{11}), where we replace $\rho_1=\rho_2\rightarrow r, \, 
g\rightarrow t, \, \phi\rightarrow\omega$: 
\bea											\label{21}
\cosh l=\cosh^2r\cosh t-\sinh^2r\cos\omega \, .
\eea 
For fixed $t$ and $\omega$ this gives $r(l)$, from which we derive $dr/dl$. 
Since neither the volume $\VgB$ nor the areas $\SgBr$ depend on $\omega$ 
in this special screw motion of $\B$, the density $\QgBr$ is again given by 
(\ref{19}). 
The density $\PgBl$ is then 
\bea											\label{22}
\PgBl=\frac{2\pi\sinh l}{\VgB(\cosh t - \cos\omega)}\left[\cosh^{-1}\left(\cosh a 
\sqrt{\frac{\cosh t-\cos\omega}{\cosh l-\cos\omega}}\right)-\frac{t}{2}\right] \, ,		
\eea
with $\VgB$ as given in (\ref{17}) with $M=t/2$ . 
A few sample plots of $\PgBl$ are given in figure 7. 

\vskip3mm  \hskip3cm\scalebox{0.45}{\ig{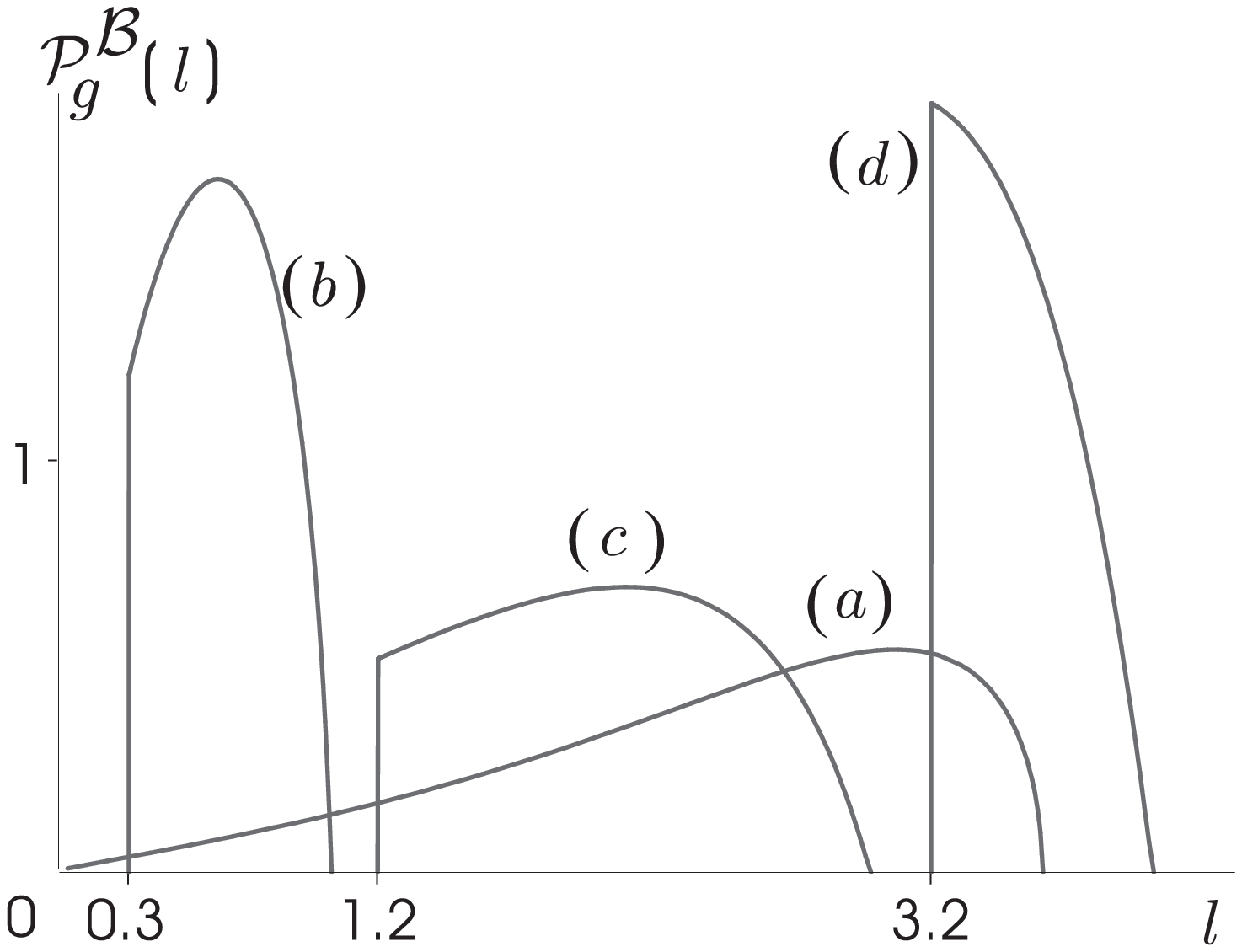}}  \vskip3mm          
    
\noindent {\small {\bf Figure 7} Probability densities $\PgBl$ of pair separations 
for special screw motions $g=(t,\omega)$ of a ball $\B$ with radius $a$, 
when the axis of the isometry crosses the centre of the ball. 
In $(a)$ we took $t=0$ (no translation), $a=1.8$, and $\omega=\pi$; 
in $(b)$, $t=0.3$, $a=1.5$, and $\omega=\pi/8$;
in $(c)$, $t=1.2$, $a=2.0$, and $\omega=\pi/8$; 
and in $(d)$, $t=3.2$, $a=2.0$, and $\omega=\pi$. 
Each plot starts at $l=t$, 
and abruptly except when we have pure rotation $(t=0,$ case $(a))$.
All plots end when 
$\cosh l=\cosh^2\rho\cosh t - \sinh^2\rho\cos\omega$, 
with $\cosh\rho=\cosh a$ sech$(t/2)$.
All integrated areas are unitary. }
\vskip5mm 

We clearly recover the equation (\ref{20}) when $\omega=0$.  
On the other hand, setting $t=0$ in (\ref{22}) gives the $\PgBl$ for a special   
elliptic isometry, namely a pure rotation $\omega$ of the ball $\B$   
when the axis of the rotation contains the centre of the ball:   
\bea											\label{23}
\PgBl=\frac{\pi\sinh l}{\VgB\sin^2(\omega/2)}\cosh^{-1}\left(\frac
{\cosh a \sinh(\omega/2)}{\sqrt{\sinh^2(l/2)-\sinh^2(\omega/2)}}\right)\,;
\eea
a special $\PgBl$ with $t=0$ is in figure 7$(a)$.
\section{Parabolic motions}                              
To describe a {\it parabolic} isometry $g$ of $H^3$ we need first announce its 
2-parametric apex $A$, a point at infinity.  
Next we select an arbitrary point $C$ of $H^3$, and draw the unique horosphere 
$\C$ with centre $A$ and containing $C$.  
Then, starting from $C$ we mark an arc of horocycle with length $\mu$, 
laying on $\C$; the direction of the arc and the value of $\mu$ demand  
two new parametres and finally fix the isometry $g$. 
The horocyclic separation between $C$ and its $g$-transported $C_g$ being 
$\mu$, the corresponding geodetic separation $m$ is given by eq.$(5c)$: 
\bea      										\label{24}
\mu=2\sinh(m/2) \,. 
\eea 

We now consider a solid ball $\B$ with radius $a$ and centre $C$; 
clearly there is no loss of generality in this last choice. 
The two parametres $(a, m)$ suffice to completely describe $\PgBl$. 

Denote as $r$ the geodetic altitude of a point $P$ of $H^3$ relative to 
the horosphere $\C$; 
$r$ is counted positive if $P$ is outside $\C$, and negative if inside. 
Also draw the horosphere $\Cr$ with apex $A$ and intersecting $P$. 
Under the isometry $g$ all points of $\Cr$ are equally displaced along 
horocyclic arcs laying on $\Cr$, and measuring 
\bea											\label{25} 
\lambda=\mu\, e^r \, ;
\eea 
equivalently, the geodetic separation $l$ between $P$ and $P_g$ is given by 
(see figure 8)  
 \bea											\label{26}
\sinh(l/2)=e^r \sinh(m/2) \, .
\eea 

\vskip3mm  \hskip3cm\scalebox{0.5}{\ig{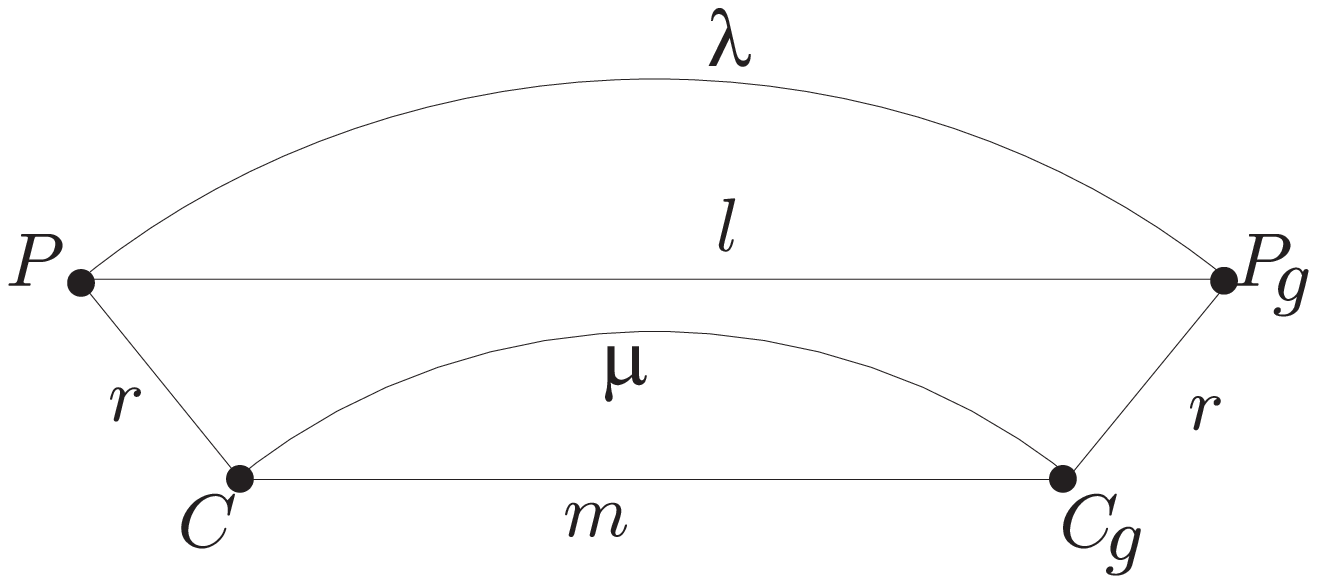}}  \vskip3mm          
    
\noindent {\small {\bf Figure 8} A parabolic isometry $g$ of $H^3$ brings the 
points $C$ and $P$ to $C_g$ and $P_g$, respectively; $m$ and $l$ are geodetic 
arcs, $\mu$ and $\lambda$ are horocyclic arcs; $r$ are parallel geodetic arcs 
orthogonal to both $\mu$ and $\lambda$; all arcs lay in a same $H^2$. }
\vskip5mm 

\noindent For future use we compute $dr/dl$ from (\ref{26}), with $m$ fixed:  
\bea											\label{27}
\frac{dr}{dl}=\frac{1}{2}\coth(\frac{l}{2})\,. 
\eea 

The horosphere $\Cr$ intersects each solid ball $\B$ and $\Bg$ in flat circular 
disks $\Dr$ and $\Drg$, both with radius $\rho$. 
To have $\rho$ as a function of $r$ and $a$ we introduce an auxiliary variable 
$s$ (see figure 9) and solve the system 
\bea											\label{28} 
\cosh a=\cosh s\cosh(r-f)\,,\hskip8mm e^f=\cosh s\, ,\hskip8mm \rho=\sinh s\, , 
\eea 
which gives 							
\bea		                                                      \label{29}
\rho=\sqrt{2(\cosh a - \cosh r)e^r} \, . 
\eea 

\vskip3mm  \hskip3cm\scalebox{0.5}{\ig{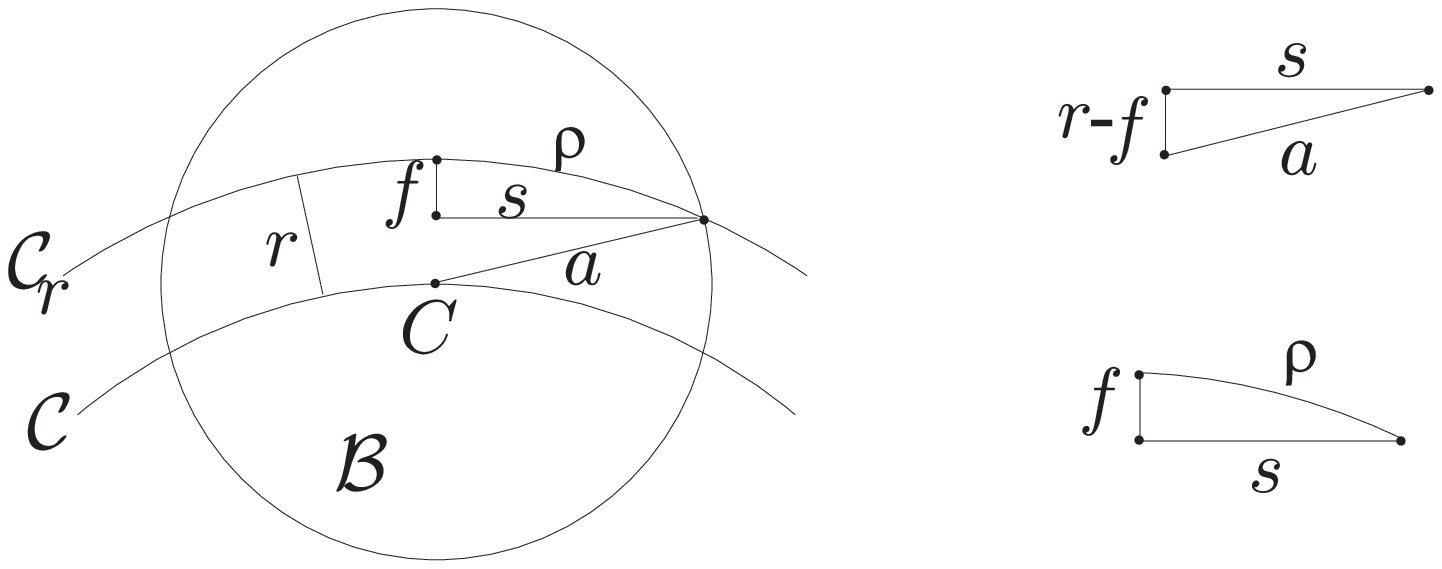}}  \vskip3mm      
    
\noindent {\small {\bf Figure 9} $\Cr$ and $\C$ are equidistant horospheres 
both intersecting the solid ball $\B$ with centre $C$ and radius $a$. 
We note a right-angled geodetic triangle with sides $a$ (hypotenuse), $s$, 
and $r-f$, where $f$ is the geodetic arc orthogonal to both 
the geodetic arc $s$ and the horocyclic arc $\rho$. }
\vskip5mm 

Our geometric situation is now phrased in the following terms: 
in a 2D flat plane (the horosphere $\Cr$) we have two circular disks $\Dr$ 
and $\Drg$, both with radius $\rho$, whose centres are separated by $\lambda$. 
We ask for the probability $\RgBr dr$ that a randomly chosen $g$-pair 
$(P, P_g)$, such that $P_g\in\BUBg$, has altitude between $r$ and $r+dr$. 
Clearly the probability density $\RgBr$ is proportional to the area $\SgBr$ 
of the intersection $\Dr\cap\Drg$, the coefficient of proportionality 
being the inverse of the volume $\VgB$ of the solid lens $\BUBg$: 
\bea											\label{30} 
\RgBr=\frac{\SgBr}{\VgB}\, . 
\eea 
The euclidean area $\SgBr$ is simple to obtain (see figure 10), it is 
\bea											\label{31}
\SgBr=4\left[\frac{1}{2}\alpha\rho^2
              -\frac{\lambda}{4}\sqrt{\rho^2-\lambda^2/4}\right] 
\eea 
with $\lambda=2\sinh(l/2)$, $\rho(r)$ as in (\ref{29}), 
and $\cos\alpha(r)=\lambda/(2\rho)$\,. 

\vskip3mm  \hskip3cm\scalebox{0.5}{\ig{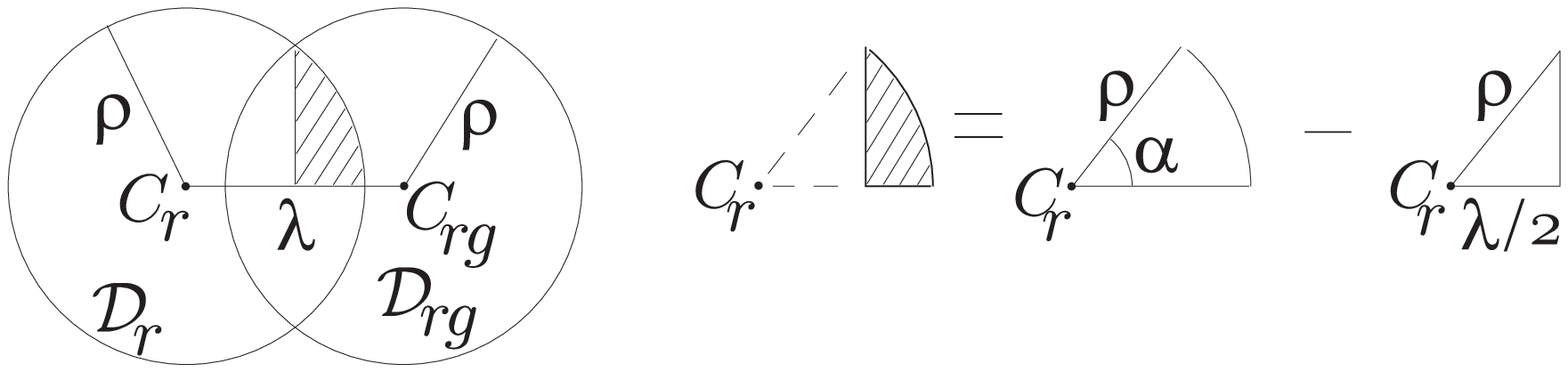}}  \vskip3mm      
    
\noindent {\small {\bf Figure 10} $\Dr$ and $\Drg$ are coplanar flat circular 
disks with radius $\rho$ and with separation $\lambda$ between their centres 
$C_r$ and $C_{rg}$. The area $\SgBr$ of the intersection $\Dr\cap\Drg$ 
is four times the shadowed area. }
\vskip5mm 

Since the probabilities $\PgBl dl$ and $\RgBr dr$ are the same, we finally have 
\bea											\label{32}
\PgBl&=&\frac{dr}{dl}\RgBrl=\frac{dr}{dl}\frac{\SgBrl}{\VgB}= \\ 
     &=&\frac{\coth(l/2)}{\VgB}                                        
\left[\rho^2\cos^{-1}\left(\frac{\sinh(l/2)}{\rho}\right)-
\sinh(l/2)\sqrt{\rho^2-\sinh^2(l/2)}\right]\, , 
\eea 
with 
\bea											\label{34} 
\rho(l)=\sqrt{\sinh^2a-\left[\cosh a -\sinh(l/2)/\sinh(m/2)\right]^2} \, . 
\eea 
See figure 11, where examples of $\PgBl$ for parabolic isometries are 
reproduced. 
In each plot we have $l_{max}$ and $l_{min}$ given respectively by 
$\sinh(l/2)=\tanh(m/2)e^{\pm R}$, with $\cosh R=\cosh a$ sech$(m/2)$.

\vskip3mm  \hskip3cm\scalebox{0.7}{\ig{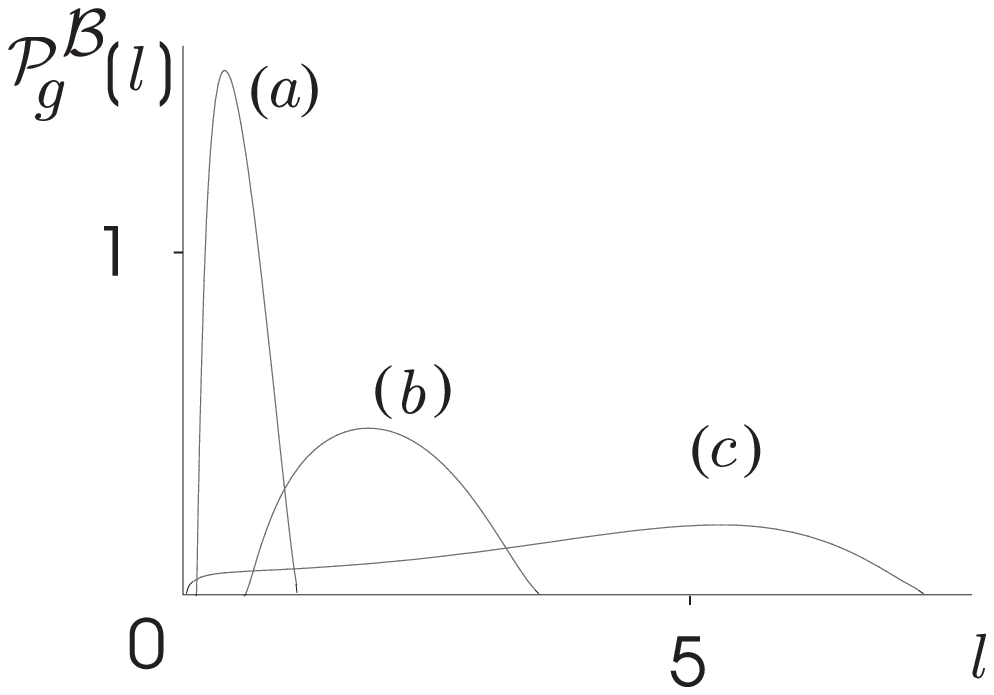}}  \vskip3mm      
    
\noindent {\small {\bf Figure 11} Probability densities $\PgBl$ of pair 
separations for parabolic isometries $g$ of a solid ball $\B$ with radius $a$. 
Under $g$, the centre of the ball is displaced a geodetic distance $m$. 
In $(a)$ we took $a=1$ and $m=0.5$; in $(b)$, $a=8$ and $m=15$; 
and in $(c)$, $a=4$ and $m=3$. All integrated areas are unitary. }
\vskip5mm 
\section{General translations}                              
We now generalize the special translations of section 5. 
We consider a hyperbolic isometry $g$ of $H^3$ whose axis is $\zeta$, 
and value $t$ measured along the axis. 
The solid ball $\B$ with radius $a$ now has centre $C$ at a distance $b$ from the
axis; in section 5 we assumed the special value $b=0$. 
Under the isometry $g$ the centre $C_g$ of the new ball $\Bg$ is separated $m$ 
from $C$; according to (\ref{6}), we have 
\bea											\label{35}
\sinh(m/2)=\sinh(t/2)\cosh b. 
\eea 

We clearly have nonempty intersection $\BUBg$ only when $m<2a$; 
values of the parametres $t, b$, and $a$ interesting for our purposes then obey 
the constraint 
\bea											\label{36}
\sinh(t/2)\cosh b < \sinh a .
\eea 
The thickness $T$, radius $R$, and volume $\VgB$ of the solid lens $\BUBg$ are 
still given by (\ref{12}), (\ref{13}), and (\ref{17}), 
with $2M=m(t, b)$ as in (\ref{35}). 

To obtain the probability density $\PgBl$ we follow the same four steps 
as described in ref.\cite{cceuc}. 
The first step is to investigate the shape of the surface $\BUCr$, 
where $\Cr$ is the infinitely long cylinder with axis $\zeta$ and radius $r$. 
The surface $\BUCr$ is either a topological annulus (if $b+r<a$), 
or a topological disk (if $a, b$, and $r$ can form a triangle), 
or is empty (if $a<|b-r|$). 
To have the dimensions of $\BUCr$ we consider a generic point 
$B=(r, \phi, \zeta)$ of its contour; we note that the distance from $B$ 
to the centre $C=(b, 0, 0)$ of $\B$ is the radius $a$, then (\ref{11}) gives 
$\zeta(a, b, r, \phi)$ according to 
\bea											\label{37} 
\cosh a=\cosh b\cosh r\cosh\zeta-\sinh b\sinh r\cos\phi ;
\eea 
the variable half width $z(\phi)$ of the intersection is then 
(eq.$(5a)$ with $e'\rightarrow z$ and $g\rightarrow\zeta$) 
\bea											\label{38}
z=\zeta\cosh r=\cosh^{-1}(\alpha+\beta\cos\phi)\cosh r , 
\eea 
where 
\bea											\label{39}
\alpha=\frac{\cosh a}{\cosh b\cosh r}, \hskip5mm \beta=\tanh b\tanh r .
\eea
For fixed values of $a, b,$ and $r$, the intersection $\BUCr$ lies between 
the curves $z(\phi)$ and $-z(\phi)$.  
Figure 12($a$) depicts an annulus-like intersection $\BUCr$, 
which occurs whenever $0<r<a-b$ (equivalently $\alpha>\beta+1$); 
note that the equator of the annulus measures 
$2\pi\sinh r$, due to the azimuthal factor $g_{\phi\phi}=\sinh^2{\rho}$ in (7). 
Clearly the extremes $-\pi$ and $\pi$ of $\phi$ are identified. 

\vskip3mm  \hskip2cm\scalebox{0.6}{\ig{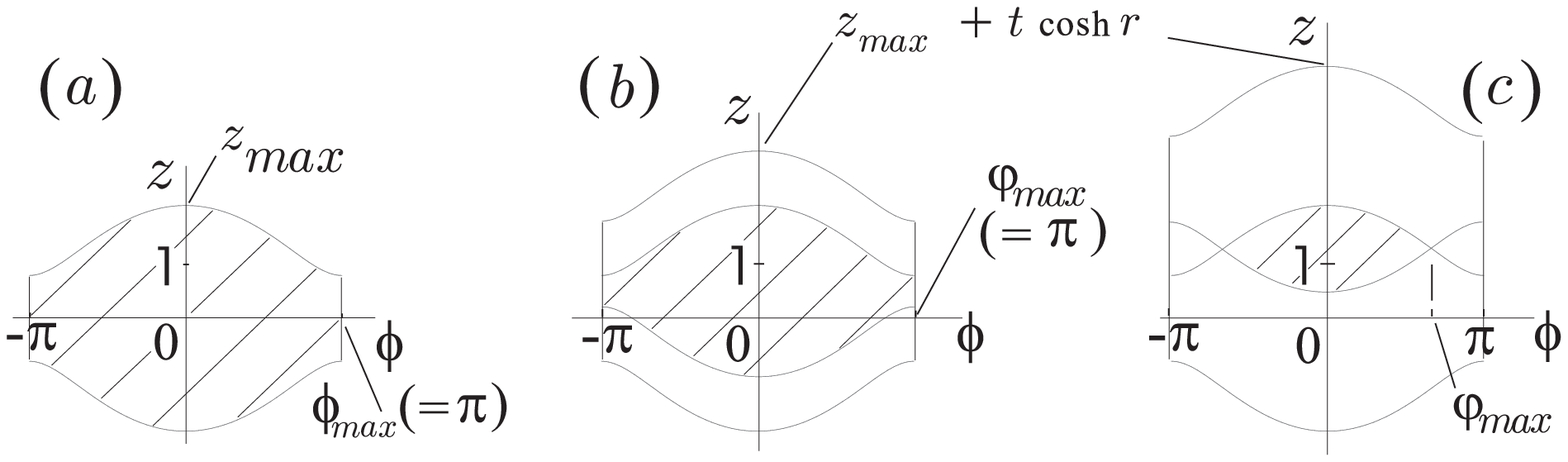}}  \vskip3mm      
    
\noindent {\small {\bf Figure 12} Sample intersections with $a=2, b=1, $ 
and $r=0.9$. \\
$(a)$ The annulus-like intersection $\BUCr$. \\
$(b)$ The (dashed) annulus-like intersection $\BUBgUCr$ when $t\cosh r=1$. \\
$(c)$ The (dashed) disk-like intersection $\BUBgUCr$ when $t\cosh r=2.5$. }
\vskip5mm 
On the other hand, figure 13($a$) shows a disk-like intersection $\BUCr$, 
which occurs whenever $\alpha<\beta+1$, 
with unequal radii $z_{max}$ and $\phi_{max}\sinh r$, with 
\bea											\label{40} 
z_{max}=\cosh^{-1}(\alpha+\beta)\cosh r \,, \hskip5mm 
\phi_{max}=\cos^{-1}\left(\frac{1-\alpha}{\beta}\right). 
\eea 

\vskip3mm  \hskip25mm\scalebox{0.5}{\ig{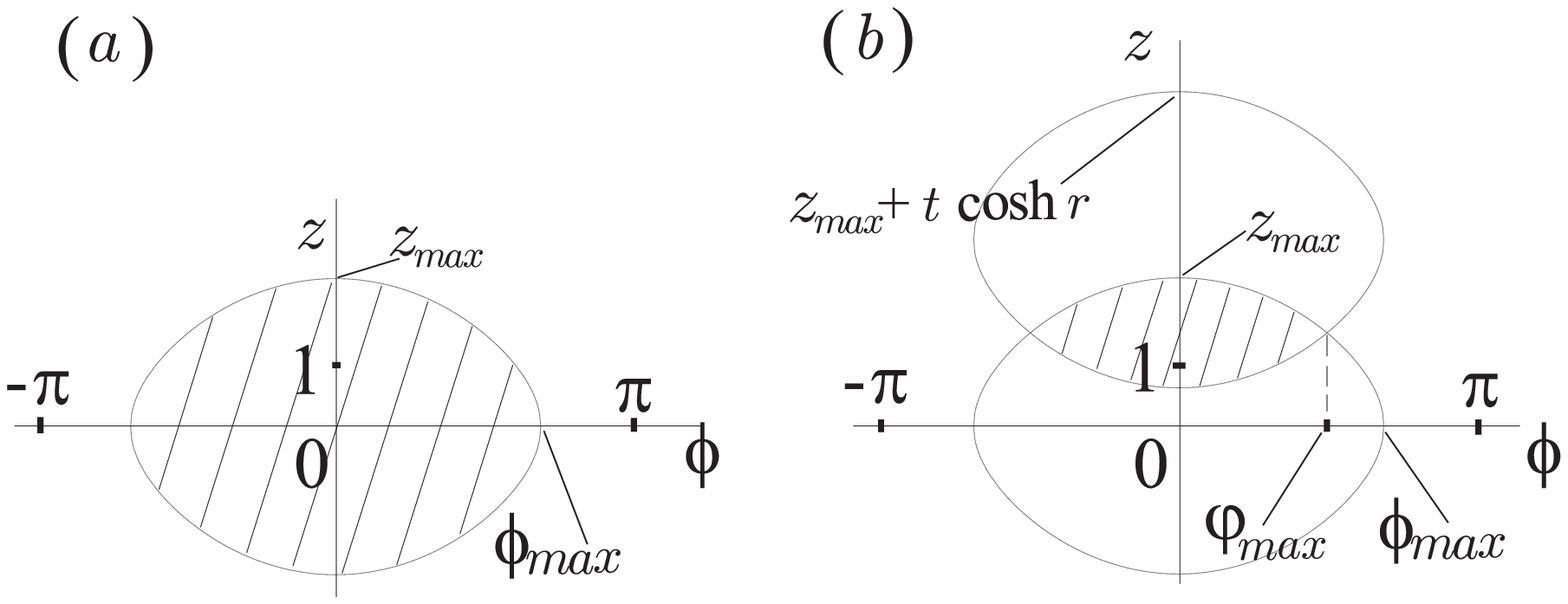}}  \vskip3mm      
    
\noindent {\small {\bf Figure 13} Sample intersections with $a=2, b=1, $ 
and $r=1.2$. \\
$(a)$ The disk-like intersection $\BUCr$. \\
$(b)$ The disk-like intersection $\BUBgUCr$ when $t\cosh r=3$.}
\vskip5mm 

The second step is to investigate the shape of the combined intersection 
$\BUBgUCr$. 
To examine the possible occurrence of annulus-like intersections $\BUBgUCr$ 
we project the centres of $\B$ and $\Bg$ on the $\zeta$ axis, 
and consider the midpoint of these projections; if this midpoint lies 
inside the solid balls, that is, if 
$\cosh a>\cosh b \cosh t/2$, then annulus-like intersections $\BUBgUCr$ 
may occur. Otherwise all intersections $\BUBgUCr$ are disk-like. 
Clearly $\BUBgUCr$ is the intersection of $\BUCr$ with $\BgUCr$, 
and $\BgUCr$ is an exact copy of $\BUCr$, 
only longitudinally displaced a horocyclic distance $t\cosh r$ along $\Cr$.  
When $\BUCr$ is annulus-like ($0<r<a-b$), then $\BUBgUCr$ is either annulus-like 
(when $\cosh(t/2)<\alpha-\beta$, see figure $12(b)$, or disk-like (when $\alpha-\beta
<\cosh(t/2)<\alpha+\beta$, see figure $12(c)$, or is empty 
(if $\cosh(t/2)>\alpha+\beta$). 
In the disk-like intersections $\BUBgUCr$ the disk extends from $-\varphi_{max}$ 
to $\varphi_{max}$, with (see figures $12(c)$ and $13(b)$ ) 
\bea											\label{41}
\varphi_{max}=\cos^{-1}\left(\frac{\cosh(t/2)-\alpha}{\beta}\right). 
\eea 
When $\BUCr$ is disk-like, then $\BUBgUCr$ is either disk-like (when 
$\cosh(t/2)<\alpha+\beta$, see figure $13(b)$ ), or is empty 
(if $\cosh(t/2)>\alpha+\beta$). 

The third step is to evaluate the area $\SgBr$ of the surface $\BUBgUCr$. 
To this end we define the auxiliary function 
\bea											\label{42}
{\mathcal A}(\alpha, \beta, \varphi_{max})=\int_0^{\varphi_{max}}
\cosh^{-1}(\alpha+\beta\cos\phi) d\phi ,
\eea 
in terms of which the areas such as in figures $12(b), 12(c)$, and $13(b)$ are 
\bea											\label{43}
\SgBr=[2{\mathcal A}(\alpha, \beta, \varphi_{max})-t\varphi_{max}]\sinh 2r .
\eea 

The fourth and last step is to compute 
\bea                                                              \label{44}  
\PgBl=\frac{dr(l)}{dl}\frac{\SgBrl}{\VgB} , 
\eea 
where $r(l)$ is found from (\ref{18}), and $\VgB$ from (\ref{17}) and (\ref{35}). 
In figure 14 we reproduce three examples of the density $\PgBl$ for 
general hyperbolic translations; clearly all plots start abruptly 
at $l_{min}\geq t$. 

\vskip3mm  \hskip1cm\scalebox{0.8}{\ig{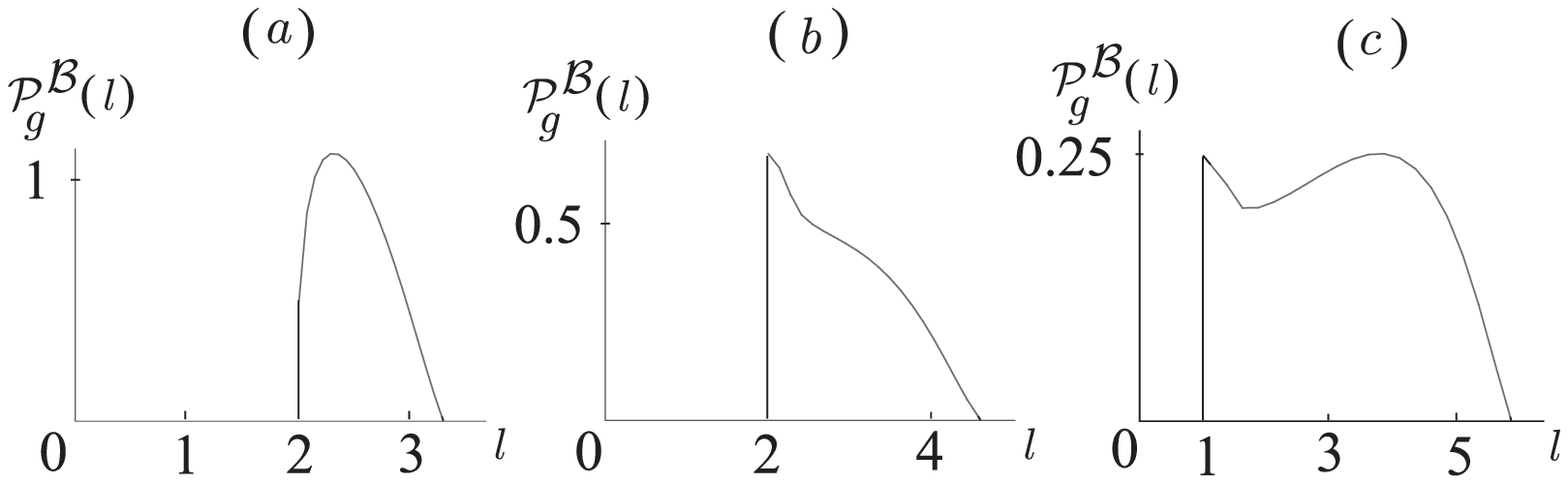}}  \vskip3mm   
    
\noindent {\small {\bf Figure 14} Sample $\PgBl$ for general translations 
of $H^3$; all integrated areas are unitary.\\
$(a)$ Here $a=3.5,\, b=3.1,$ and $t=2$; all intersections $\BUCr$ and $\BUBgUCr$ 
are disk-like; $l_{min}=t+0.002$. \\ 
$(b)$ Here $a=3.5,\, b=2.5,$ and $t=2$; there are disk-like and annulus-like 
intersections $\BUCr$, but all $\BUBgUCr$ are disk-like; $l_{min}=t$. \\ 
$(c)$ Here $a=3.5,\, b=2.5,$ and $t=1$; there are both types of intersections 
of both $\BUCr$ and $\BUBgUCr$; $l_{min}=t$.}

\section{General screw motions}                                 
We already have all elements needed to obtain the probability density 
$\PgBl$ of conjugate pair separations for a general screw motion, 
thus generalizing the results of the preceding sections 5, 6, and 8. 
We now make use of all four independent parametres, namely 
\begin{itemize}
\item $a$ = radius of the solid balls $\B$ and $\Bg$, 
\item $b$ = distance from the centres of the balls to the axis $\zeta$ 
            of the isometry, 
\item $t$ = translation of the isometry, measured along the axis, and 
\item $\omega$ = angle of rotation of the isometry, around the axis. 
\end{itemize}
We shall further write all mathematical expressions in a form appropriate for 
automatic calculation of $\PgBl$ in a computer . 
Without loss of generality for our purposes we assume $t\geq 0$ and 
$0\leq\omega\leq\pi$. 

The separation $2M$ between the centres of $\B$ and $\Bg$ is now given by (\ref{10}) 
\bea 										       \label{45} 
\cosh{2M}=\cosh^2{b}\cosh t - \sinh^2{b}\cos\omega , 
\eea 
and the condition $M<a$ is necessary to have nonempty intersection $\BUBg$. 
Assuming this condition is fulfilled, the solid lens $\BUBg$ has thickness $T$, 
radius $R$, and volume $\VgB$ given by (\ref{12}), (\ref{13}), and (\ref{17}), 
respectively, with $M$ given by (\ref{45}). 
The centre of the lens is at a distance $\sigma$ from the axis $\zeta$, 
\bea 											\label{46}
\tanh\sigma=\frac{\tanh b\cos{\omega/2}}{\cosh{t/2}}, 
\eea 
and there always exists one diametre of the lens which is directed 
perpendicular to the axis $\zeta$. 
The lens intersects the axis whenever $\sigma<R$, or equivalently 
$\cosh b\cosh t/2<\cosh a$. 
 
We next imagine an infinite  family of sufficiently long, coaxial 
(axis $\zeta$),  cylindrical surfaces $\Cr$ with variable radius $r$. 
We are interested in the intersection of each $\Cr$ with the solid lens $\BUBg$; 
clearly only values of $r$ in the range $(r_{min}, r_{max})$ 
give nonempty intersections $\BUBgUCr$, where  
\bea											\label{47} 
r_{min}=(\sigma-R)\,\Theta(\sigma-R)\, , \hskip5mm r_{max}=\sigma+R .
\eea 
Here $\Theta$ is the step function with values 0 and 1. 
Our strategy to approach $\BUBgUCr$ is first study $\BUCr$, then $\BgUCr$, 
and finally the intersection of these two. 

For a given $r\in(r_{min}, r_{max})$ we note that $\BUCr$ is annulus-like 
when $r<a-b$, and disk-like otherwise, so we define 
\bea 											\label{48} 
\phi_{max}=\pi\,\Theta(a-b-r)+
\cos^{-1}\left(\frac{1-\alpha}{\beta}\right)\Theta(r+b-a) \, , 
\eea 
where $\alpha(a, b, r)$ and $\beta(b, r)$ were given in (\ref{39}). 
Each intersection $\BUCr$ is nonempty for 
$\phi\in(-\phi_{max}, \phi_{max})$, and lies between the two curves 
\bea 											\label{49} 
z_1=\cosh^{-1}(\alpha+\beta\cos\phi)\cosh{r}\,\Theta(\phi+\phi_{max})\,
\Theta(\phi_{max}-\phi),  \hskip5mm z_2=-z_1 ; 
\eea 
these two curves are drawn on the geometrically flat cylinder $\Cr$. 

The intersection $\BgUCr$ is identical to $\BUCr$, but is displaced $t\cosh r$ 
longitudinally on $\Cr$, and $\omega$ azimuthally. 
It thus lies between the curves 
\bea											\label{50}
z_3=\tau+\cosh^{-1}\left[\alpha + \beta\cos(\phi-\omega)\right]\cosh{r} 
\nonumber \\ \times [ \Theta(\phi-\omega + \phi_{max})\,\Theta(\phi_{max}-\phi) + 
\Theta(\phi_{max}-\phi+\omega-2\pi)\,\Theta(\phi+\phi_{max}) ],  
\eea
\bea           									\label{51}
z_4=-z_3+2\tau , 	     			   	                  
\eea 
where 
\bea											\label{52}
\tau=t\cosh{r}\,\Theta(\phi+\phi_{max})\,\Theta(\phi_{max}-\phi) ; 
\eea
the term with the $\Theta$ function containing $2\pi$ in eq. (\ref{50}) 
is included to allow automatic computing. 

As in ref.\cite{cceuc}, the area $\SgBr$ of the intersection $\BUBgUCr$ is 
\bea 											\label{53}
\SgBr =
\sinh{r} \int_{-\phi_{max}}^{\phi_{max}}\Theta(z_1-z_4) 
\left[\min(z_1,z_3)-\max(z_2,z_4)\right]d\phi  .
\eea
Finally, the probability density $\PgBl$ is given by (\ref{44}) 
with $r(l)$ coming from (\ref{21}); we find that 
\bea											\label{54}
\frac{dr}{dl}=\frac{\sinh{l} \, {\rm csch} 2r}{\cosh{t}-\cos\omega}. 
\eea
In figure 15 a few sample plots of $\PgBl$ for screw motions in $H^3$ are given. 
 
\vskip3mm  \hskip4cm\scalebox{0.9}{\ig{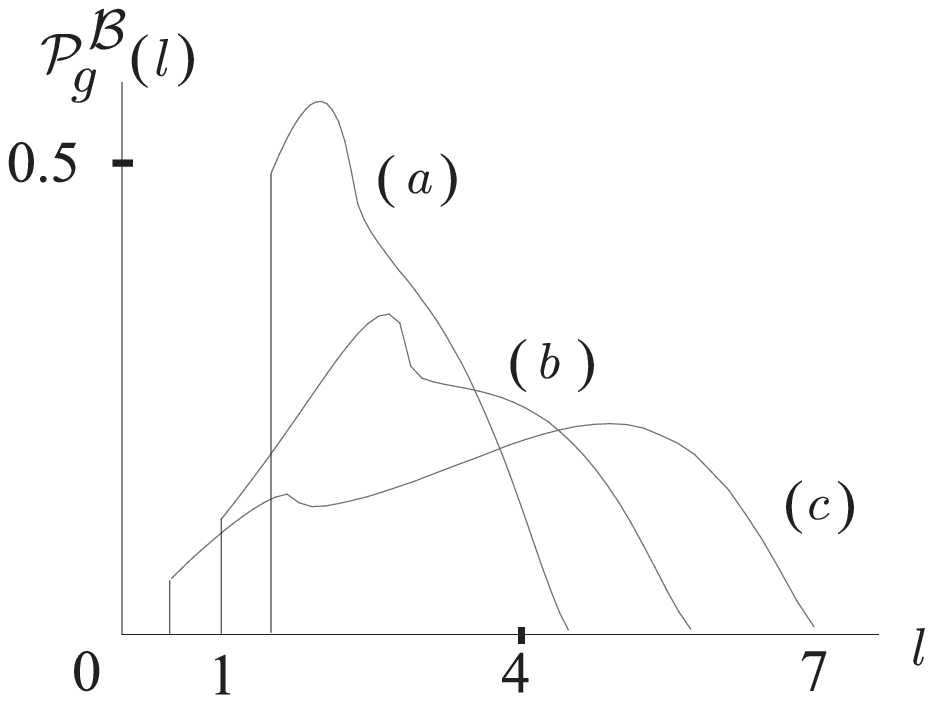}}  \vskip3mm   
    
\noindent {\small {\bf Figure 15} Sample $\PgBl$ for general screw motions  
of $H^3$; all integrated areas are unitary.\\
$(a)$ Here $a=4,\, b=3,\, t=1.5$, and $\omega=\pi/2$.  \\ 
$(b)$ Here $a=4,\, b=2.5,\, t=1$, and $\omega=\pi/2$.  \\ 
$(c)$ Here $a=4,\, b=2.5,\, t=0.5$, and $\omega=\pi/4$.  }
 
\section{Conclusion}                                             
The three plots in figure 15 are the output of a computer program whose 
inputs are the values of $a, b, t$, and $\omega$; 
given these inputs, the program proceeds without any intervention. 
 
The three parametres $(t, \omega, b)$ related to the screw motion can be  
extracted from a $4\times 4$ matrix $\M_g$, which expresses the motion 
in terms of the minkowskian coordinates $(W; X, Y, Z)$ \cite{SnapPea}. 
Indeed, it can be shown that the trace $T$, the sum 
$\Sigma$ of the principal minors of order 2, and the time-time coefficient $U$ 
of the matrix $\M_g$ are  
\bea									          \label{55}
T=2(\cosh t + \cos\omega), \hskip3mm \Sigma=2(1+2\cosh t\cos\omega), 
\hskip3mm U=\cosh t\cosh^2b-\cos\omega\sinh^2b ;
\eea 
if we define $V=\sqrt{8+T^2-4\Sigma}$ \, then we obtain 
\bea											\label{56}
\cosh t=\frac{1}{4}(T+V), \hskip3mm 
\cos\omega=\frac{1}{4}(T-V), \hskip3mm 
\cosh 2b=\frac{4U-T}{V}. 
\eea

To close this report we show through a concrete example how to use the 
functions $\PgBl$ to get information about the topology of the universe.  
We need first briefly recall the theory that underlies the subject; 
for details see \cite{spikes1}, \cite{sccII}, \cite{tsct}. 

Expected (or theoretical) normalized histograms $\phi(i)$ of pair separations 
are decomposed as 
\bea                           						\label{57}
\phi(i)=\phi^{un}(i)+\frac{1}{n-1} \sum_g\nu_g\left[\phi^g(i)-\phi^{un}(i)\right],
\eea 
where $\phi^{un}(i)$ is the expected normalized histogram of the uncorrelated 
pair separations, 
the $i$ denotes an interval of separations (a bin in the histogram), 
$n$ is the finite number of objects in the solid ball $\B$, 
$\nu_g=N_g/n=\nu_{g^{-1}}$ with $N_g$ the number of $g$-pairs with both 
members inside $\B$, and 
$\phi^g(i)$ is the expected normalized histogram of the $g$-pairs. 
These $\phi^g(i)$ are the histogramic counterparts of the functions 
$\PgBl$ of \cite{cceuc} and of this report. 
From (\ref{57}), and accepting a suggestion by Fagundes and Gausmann 
\cite{fagaus}, we write 
\bea											\label{58} 
(n-1)[\phi-\phi^{sc}]=
(n-1-\sum\nu_g)[\phi^{un}-\phi^{sc}]+\sum\nu_g[\phi^g-\phi^{sc}], 
\eea 
where $\phi^{sc}(i)$ is the expected normalized histogram of pair separations 
in a simply connected ball with same radius and geometry as $\B$; 
it is the histogramic counterpart of the function  
${\mathcal F}_H(a,l)$ of \cite{bt1} and of ${\mathcal F}_H(a,s)$ of \cite{mjr1}. 
In the limit $n\rightarrow\infty$ the products 
$n[\phi-\phi^{sc}]=:\varphi^{\mathcal B}$ and 
$n[\phi^{un}-\phi^{sc}]=:\varphi_{un}^{\mathcal B}$ remain finite, and we write 
\bea											\label{59} 
\varphi^{\B}(l)=\vfuBl+\vfGBl, \hskip8mm 
\vfGBl:=\sum\nu_g[\PgBl-\PscBl]. 
\eea 
To go from the various histograms $f(i)$ to the corresponding functions $f(l)$ 
we have simply made the number of bins tend to infinity. 
The function $\varphi^{\B}(l)$ has been called the topological signature 
of a ball $\B$ in a multiply connected space; 
since in practice the function $\vfuBl$ is usually small valued 
when compared with both $\varphi^{\B}(l)$ and $\vfGBl$, 
the function $\vfGBl$ is generally a good approximation of the 
topological signature $\varphi^{\B}(l)$. 

We now turn to a specific example: that of a ball $\B$ in the Seifert-Weber 
dodecahedral space. This multiply connected hyperbolic three-space is obtained 
from a regular solid dodecahedron ${\D}$ by pairwise identifying opposite 
faces using twists of 3/10 of a revolution \cite{Weeks}. 
We make the centres of $\B$ and $\D$ coincide (a rather uncopernic assumption),  
and choose $\B$ tangent to the edges of $\D$. 

Computer simulations of $(n-1)[\phi(i)-\phi^{sc}(i)]$ for the hyperbolic 
ball $\B$ are given in the literature (see figure 7 in \cite{tsct} or 
figure 3 in \cite{ssu}), and we now construct its {\it approximate} expected 
counterpart $\vfGBl$, eq.(\ref{59}). 

We first select two of the 12 matrices of face-pairing isometries of $\D$: 
\bea 											\label{60} 
\M_1= \left( \begin{array}{cccc}
3.736068 & 0 & 0 & 3.599751 \\ 
0 & -0.309017 & -0.951057 & 0 \\ 
0 & 0.951057 & -0.309017 & 0 \\ 
3.599751 & 0 & 0 & 3.736068 \end{array} \right) \, , 
\eea
\bea 											\label{61} 
\M_2= \left( \begin{array}{cccc}
3.736068 & -3.219715 & 0 & -1.609857 \\ 
-3.219715 & 2.927051 & 0.425325 & 1.618034 \\ 
0 & -0.425325 & -0.309017 & 0.850651 \\ 
-1.609857 & 1.618034 & -0.850651 & 0.5 \end{array} \right) \, , 
\eea
sufficient for our purposes. 
Applying (\ref{56}) to any of $\M_1$ or $\M_2$ we find the values 
\bea 											\label{62}
t_1=1.992, \hskip3mm \omega_1=108^o, \hskip3mm b_1=0, 
\eea 
for the translation, the rotation, and the distance from the axis of the 
isometry to the centre of $\B$ (the origin of coordinates). 
From (\ref{45}) and (\ref{62}) we obtain half separation $M_1=0.996$ between 
the centres of $\B$ and $\Bg$, a value smaller than the radius $a=1.439$ of $\B$. 

We also need consider 60 other isometries, whose common prototype matrix is 
\bea 											\label{63}
\M_3= \M_1\M_2= \left( \begin{array}{cccc}
8.163119 & -6.204554 & -3.062131 & -4.214661 \\ 
0.994947 & -0.5 & 0.162460 & -1.309017 \\ 
-3.062131 & 2.915224 & 0.5 & 1.275976 \\ 
7.434376 & -5.545085 & -3.178089 & -3.927051 \end{array} \right) \, ; 
\eea
each such isometry gives, from (\ref{56}), 
\bea 											\label{64}
t_3=1.746, \hskip3mm \omega_3\approx 147^o, \hskip3mm b_3=0.999.  
\eea
These 60 isometries also contribute to (\ref{59}), since from (\ref{45}) 
and (\ref{64}) we find half separation $M_3=1.395$, a value smaller than $a$. 
All other isometries seem to give $M>a$, so they do not contribute to (\ref{59}). 

For the 12 fundamental isometries (\ref{62}) we find intersections $\BUBg$ with volume  
$V_1^{\mathcal B}=1.377$ as given by (\ref{17}); 
all produce the same spectrum ${\mathcal P}_1^{\mathcal B}(l)$, 
which has nonzero values only in the interval $l\in(1.99 , 2.79)$. 
On the other hand, for the 60 isometries (\ref{64}) we find 
$V_3^{\mathcal B}=0.011433$, and a spectrum  ${\mathcal P}_3^{\mathcal B}(l)$ 
with nonzero values only when $l\in(1.75 , 1.99)$. 
Finally, the probability density $\PscBl$ for pair separations in a hyperbolic 
ball is the function ${\mathcal F}_H(a,l)$ of \cite{bt1}, or the function 
${\mathcal F}_H(a,s)$ of \cite{mjr1}; 
for unitary curvature of the space, the volume of the ball with radius $a=1.439$ 
is $V_{sc}^{\B}=18.8$. 
From (\ref{59}) we then have (see figure 16)
\bea											\label{65} 
\vfGBl=\frac{12V_1^{\B}}{V_{sc}^{\B}}
\left[ \PdBl - \PscBl \right] + 
\frac{60V_3^{\B}}{V_{sc}^{\B}}\left[ \PsBl - \PscBl \right]. 
\eea 
 
\vskip3mm  \hskip3cm\scalebox{0.4}{\ig{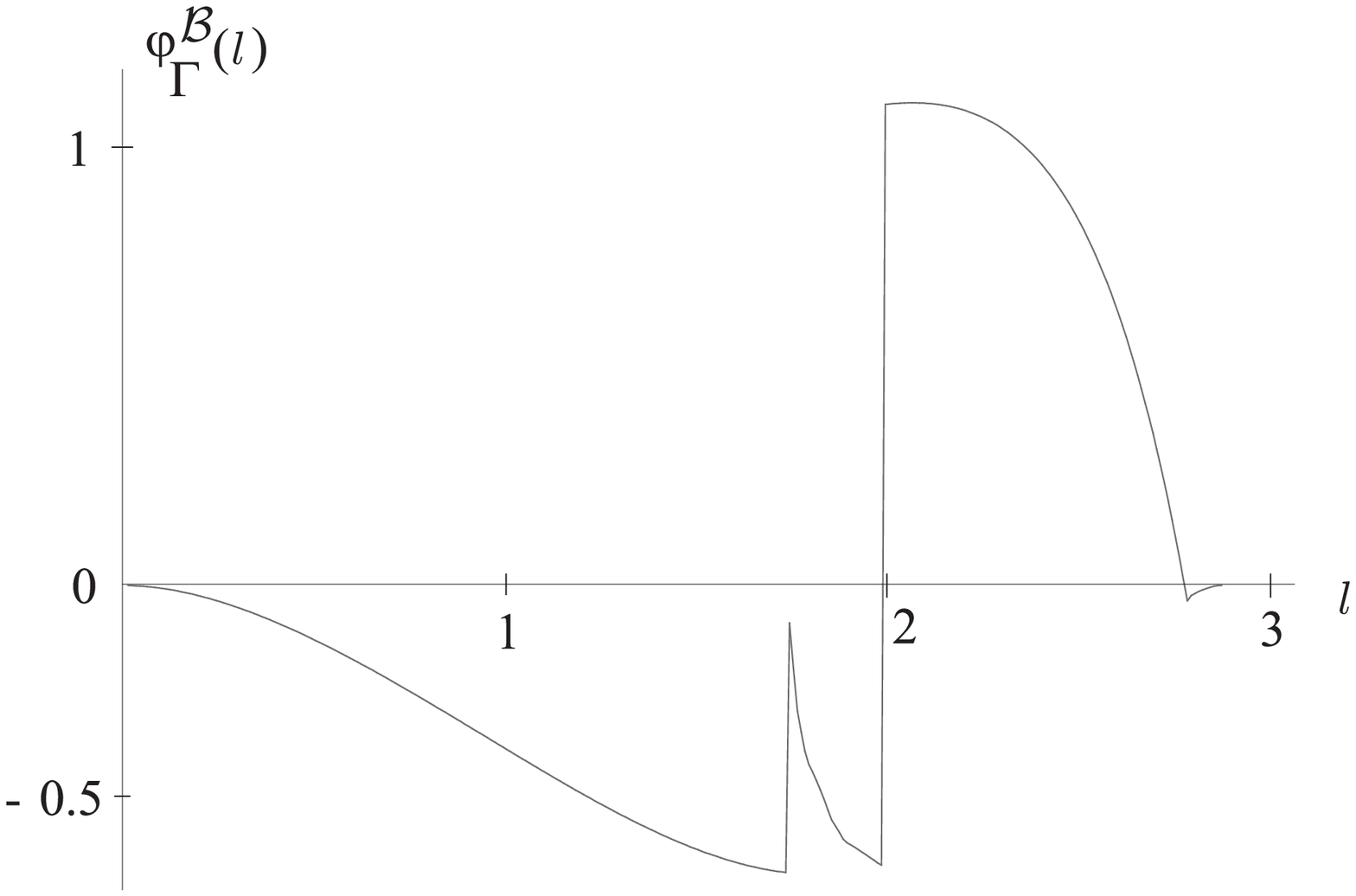}}  \vskip3mm   
    
\noindent {\small {\bf Figure 16} Approximate topological signature 
$\vfGBl$ for an observed universe endowed with the 
Seifert-Weber dodecahedral topology and unitary negative curvature. 
The centre of observation and that of the dodecahedron coincide, 
and the event horizon is supposed $a=1.44$ away. 
The discontinuities observed at $l=1.75$ and $l=1.99$ derive from isometries, 
as described in the text. Their localization and strength are good 
indicators of the topology of the universe.}
\vskip5mm

The approximate signature $\vfGBl$ of figure 16 
bears close similarity with the corresponding histograms 
figure 7 in \cite{tsct} and figure 3 in \cite{ssu}. 
However, some small distortion can be seen, probably arising because 
the uncorrelated contribution $\vfuBl$, present in (\ref{59}), 
was not taken into account. 
As a matter of fact, we have not been able to obtain the expected 
$\vfuBl$ neither for the present Seifert-Weber space  
nor for any simpler $3D$ nontrivial manifold, such as the three-torus.  
Even for the two-torus that function has been eluding our efforts; 
only for the one-torus (a circle) we have already succeeded in finding 
the $\vfuBl$ \cite{ccs1}.

\vskip1cm \noindent {\bf \Large Acknowledgments}   \\      
The author is grateful to A. Bernui, G.I. Gomero and M.J. Rebou\c{c}as 
for fruitful conversations.

\end{document}